\documentclass[aps,prb,preprint,showkeys,floatfix]{revtex4}

\usepackage{graphicx,color}
\usepackage{microtype}
\usepackage{multirow,slashbox}
\usepackage{natbib}
\usepackage[utf8]{inputenc}
\usepackage{float}
\graphicspath{{figs/}}
\usepackage{amsmath}
\usepackage{tabularx}
\usepackage[colorlinks,linkcolor=black,anchorcolor=black,citecolor=black]{hyperref}
\usepackage{amssymb}
\usepackage{subfigure}
\date{\today}

\begin{document}

\title{Misfit strain effect on the thermal expansion coefficient of graphene/MoS$_2$ van der Waals heterostructures}

\author{Run-Sen Zhang}
    \affiliation{Shanghai Key Laboratory of Mechanics in Energy Engineering, Shanghai Institute of Applied Mathematics and Mechanics, School of Mechanics and Engineering Science, Shanghai University, Shanghai 200072, People's Republic of China}

\author{Jin-Wu Jiang}
    \altaffiliation{Corresponding author: jwjiang5918@hotmail.com}
    \affiliation{Shanghai Key Laboratory of Mechanics in Energy Engineering, Shanghai Institute of Applied Mathematics and Mechanics, School of Mechanics and Engineering Science, Shanghai University, Shanghai 200072, People's Republic of China}

\date{\today}
\begin{abstract}

Van der Waals heterostructures such as graphene/MoS$_2$ are promising candidates for plenty of optical or electronic applications, owing to advanced properties inherited from the constitutional atomic layers. Thermal expansion is an important phenomenon to be considered for the thermal stability of the van der Waals heterstructure as temperature commonly rises during the operation of nano devices. In the present work, the thermal expansion coefficient for the graphene/MoS$_2$ heterostructure is investigated by molecular dynamics simulations, and the effect from the unavoidable misfit strain on the thermal expansion coefficient is revealed. The misfit strain can tune the thermal expansion coefficient by a factor of six, and this effect is quite robust in sense that it is not sensitive to the size or direction of the heterostructure. An analytic formula is derived to directly relate the thermal expansion coefficient to the misfit strain of the heterostructure, which qualitatively agrees with the numerical results although the analytic formula underestimates the misfit strain effect. Further analysis discloses that the misfit strain can efficiently engineer the thermal induced ripples, which serves as the key mechanism for the misfit strain effect on the thermal expansion coefficient. These findings provide valuable information for the thermal stability of van der Waals heterostructures and shall be benefit for practical applications of van der Waals heterostructure based nano devices.

\end{abstract}
\maketitle

\section{Introduction}

As a promising candidate for the next generation electronics, graphene has attracted lots of attention due to its outstanding thermal, mechanical and electronic properites.\cite{Geim2007The,Balandin2011Thermal,Lee2008Measurement} The absence of a finite band gap is a major shortcoming for graphene based devices. Forming van der Waals heterostructures is a very useful method to overcome the disadvantage of each individual layer and to finely tune various properties for 2D materials.\cite{Liu2016Van,Sobhit2018Structural} The combination of graphene with semiconducting two-dimensional transition metal dichalcogenides has potential applications in the field of high power electronics and novel nanometer scale devices.\cite{Balandin2008Superior,Yu2016Unusually,Zhong2012Graphene} In particular, the graphene/MoS$_2$ van der Waals heterostructure has been suggested to be used in high-performance photo detection,\cite{2017Printable} photo catalytic, molecular sieves, and electrodes.\cite{RiccardoFrisenda2018Recent,Chen2020Negative,Wu2021MoS2}  

For practical applications of the graphene/MoS$_2$ van der Waals heterosructure, it is crucial to learn the thermal expansion property of such devices because of the reduced dimensionality and high density of devices in tightly packed structures.\cite{Wang2012Electronics,Britnell2012field} The thermal expansion phenomenon is characterized by the thermal expansion coefficient (TEC). The TEC of single layer graphene and single layer MoS$_2$ have been extensively studied by both numerical simulations\cite{Mounet2005First,2009Finite,Pozzo2011Thermal,Mann2017Negative,anees2017Delineating,2014Correlation} and experiments\cite{Ortega2017Tailoring,Yoon2011Negative,Mcquade2021the,Zhang2019Thermal,Anemone2018Experimental,Lin2021Thermal}. Mounet and Marzari predicted that the TEC of single layer graphene is negative up to 2300~K.\cite{Mounet2005First} Zakhrchenko et al. showed that graphene has a negative TEC for temperatures bellow 900~K.\cite{2009Finite} It has been shown that the out-of-plane bending (flexural) mode makes a negative contribution to the TEC, while the in-plane anharmonicity makes positive contribution to the TEC.\cite{2015Theory,anees2017Delineating} The competition between these two effects leads to the negative TEC at low temperature and positive TEC at high temperature. Experiments confirmed that the TEC is negative for graphene around room temperature.\cite{bao2009controlled,Yoon2011Negative} Different from graphene, the monolayer MoS$_2$ has positive TEC. For instance, micro-Raman spectroscopy measurement obtained the in-plane TEC of monolayer MoS$_2$ as $7.6\pm 0.9\times 10^{-6}$~{K$^{-1}$} and $7.4\pm 0.5\times 10^{-6}$~{K$^{-1}$} along different directions.\cite{Zhang2019Thermal} The MoS$_2$ atomic layer has larger bending modulus than graphene, so the negative effect from the flexural mode is weaker, leading to positive TEC in the MoS$_2$ atomic layer.

As a characteristic structural feature, the constitutional layers in the van der Waals heterostructure usually have different lattice constants, which results in the misfit strain between neighboring atomic layers.\cite{Jiang2014Mechanical,Lin2014Atomically,Linyang2014Structures} The misfit strain in van der Waals heterostructures can be adjusted through various methods, including the substrate,\cite{Ni2008Raman} the structural design,\cite{Chaste2018Nanomechanical} and the photoelectrochemical etching process.\cite{shivaraman2013Raman} It has been shown that the misfit strain can cause direct effects on physical and mechanical properties of the van der Waals heterostructure.\cite{He2019Misfit}

Although the thermal expansion behavior of graphene and MoS$_2$ layers have been extensively studied, the TEC for the graphene/MoS$_2$ van der Waals heterostructure has not been well studied. Particularly, the effect of the characteristic misfit strain on the TEC is still unclear, which will be the focus of the present work.

In this paper, we perform molecular dynamics (MD) simulations to comparatively investigate the TEC for graphene/graphene bilayer, MoS$_2$/MoS$_2$ bilayer, and the graphene/MoS$_2$ heterostructure. We focus on the effect of the misfit strain on the TEC in graphene/MoS$_2$ heterostructures. We find that the TEC along the misfit strain direction decreases from $6\times 10^{-6}$~{K$^{-1}$} to $1\times 10^{-6}$~{K$^{-1}$} with misfit strain increasing from -6\% to 6\%. The misfit strain effect can be qualitatively discussed based on an analytic relation between the TEC and the misfit strain. Furthermore, we find that the thermal induced ripples are effectively tuned by the misfit strain, which can cause strong effects on the TEC of the graphene/MoS$_2$ heterostructure. The nonlinear effect also plays a role in the misfit strain effect on the TEC.

\section{Structure and simulation details}

\begin{figure}[htbp]  
  \begin{center}
    \scalebox{1.0}[1.0]{\includegraphics[width=\textwidth]{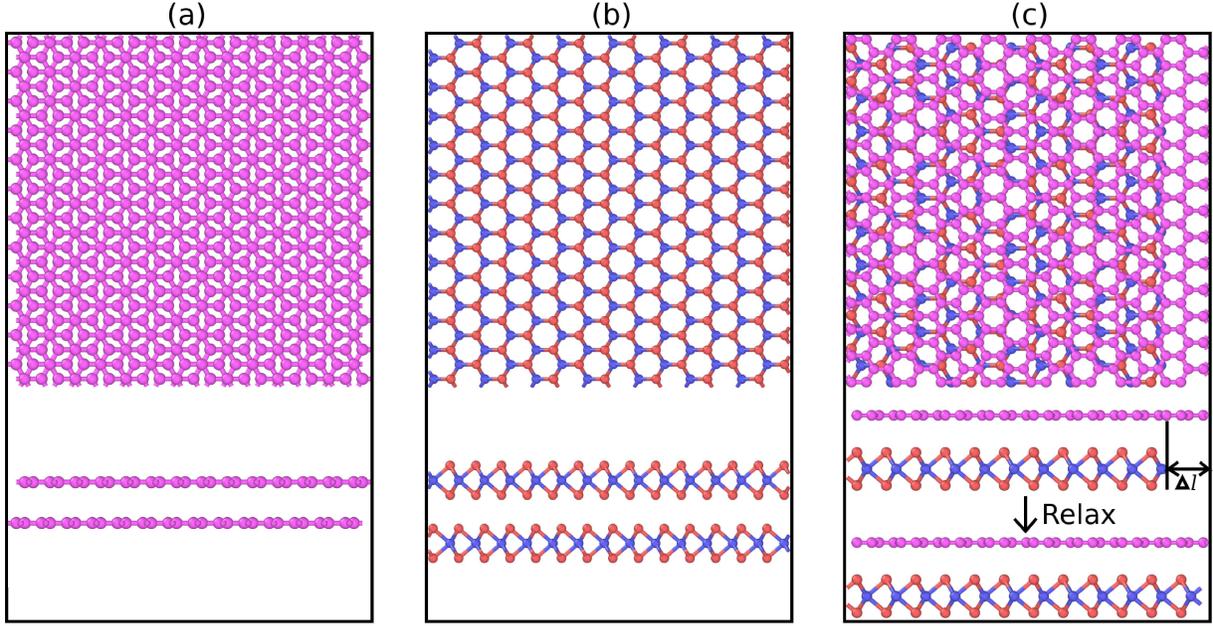}}
  \end{center}
  \caption{Bilayer structures studied in the present work. (a) The graphene/graphene structure. (b) The MoS$_2$/MoS$_2$ structure. (c) The graphene/MoS$_2$ heterostructure. The graphene and MoS$_2$ have a difference of $\Delta l$ in their length, resulting in the misfit strain in the relaxed configuration of the graphene/MoS$_2$ heterostructure.}
  \label{fig_str}
\end{figure}

Figure~\ref{fig_str} shows the three structures studied in the present work, i.e., graphene/graphene bilayer, MoS$_2$/MoS$_2$ bilayer, and graphene/MoS$_2$ heterostructure. The grahene/graphene bilayer structure is in the AB stacking order, while the MoS$_2$/MoS$_2$ structure is in its lowest-energy stacking order. The horizontal x-direction is along the armchair direction while the vertical y-direction is along the zigzag direction. The lengths of graphene and MoS$_2$ are slightly different in the graphene/MoS$_2$ heterostructure, resulting in the misfit strain as follows,
\begin{eqnarray}
\epsilon_m = - \frac{l_1 - l_2}{l_1}.
\label{eq_misfit}
\end{eqnarray}
where $l_1$ and $l_2$ are the length for the two constituting layers. The present work focuses on the effect of the misfit strain $\epsilon_m$ on the thermal expansion coefficient of the graphene/MoS$_2$ heterostructure.

The carbon-carbon interaction is described by the second generation Brenner (REBO-II) potential,\cite{Brenner2002A} while the MoS$_2$ interatomic interactions are described by the Stillinger-Weber potential\cite{StillingerF} with parameters from the recent work.\cite{jiang2019misfit} The inter-layer interactions between graphene/graphene, MoS$_2$/MoS$_2$ and graphene/MoS$_2$ are described by the Lennard-Jones potential. The distance and energy parameters in the Lennard-Jones potential are listed in table~\ref{tab_lj}.
\begin{table}[htbp]
\caption{Lennard-Jones parameters used in the present work for graphene/graphene,\cite{Jiang2015A} MoS$_2$/MoS$_2$,\cite{Jiang2015A} and graphene/MoS$_2$.\cite{Jiang2014Mechanical} }
\label{tab_lj}
\begin{tabular}{@{\extracolsep{\fill}}|c|c|c|c|}
\hline 
  structures  & $\epsilon$~{(meV)} & $\sigma$~({\AA}) & cut off~({\AA})\tabularnewline
\hline 
\hline 
graphene/graphene & 2.96 & 3.382 & 10.0\tabularnewline
\hline
MoS$_2$/MoS$_2$ & 23.6 & 3.18 & 10.0\tabularnewline
\hline
graphene/MoS$_2$ & 3.95 & 3.625 & 10.0\tabularnewline
\hline 
\end{tabular}
\end{table}

All MD simulations are performed based on the LAMMPS package,\cite{plimpton1995fast} and the OVITO package is used for visualization.\cite{stukowski2009visualization} The standard Newton equations of motion are integrated in time using the velocity Verlet algorithm with a time step of 1~{fs}. Periodic boundary conditions are employed in the two in-plane directions, while the free boundary condition is applied in the out-of-plane direction.

The thermal expansion phenomenon is simulated as follows. Firstly, the structure is minimized to the lowest-energy configuration using the conjugate gradient algorithm and then optimized for 200~{ps} within the NPT ensemble (i.e. the particles number N, the pressure P and the temperature T of the system are constant) at desired temperatures at zero pressure. Secondly, the structure is allowed to evolve for 1~{ns} within the NPT ensemble, and the size for the structure is averaged during this step. The averaged size $L$ at different temperature $T$ is then used to compute the thermal expansion coefficient. The standard deviation error $\Delta L$ in the time average gives the error bar for the size.

\section{Results and discussion}
\subsection{Simulation results}

\begin{figure}[htbp]  
  \begin{center}
    \scalebox{1.0}[1.0]{\includegraphics[scale=0.7]{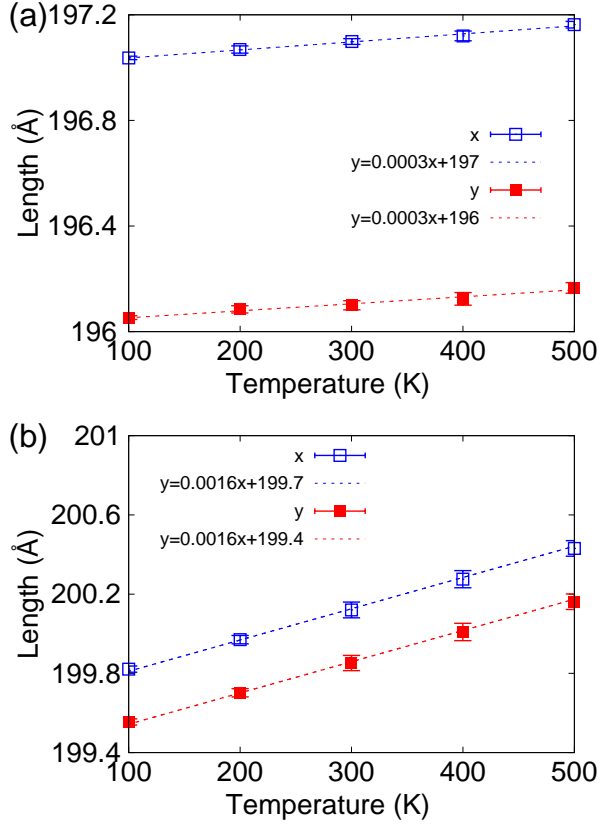}}
  \end{center}
  \caption{The temperature dependence for the length of (a) the graphene/graphene structure and (b) the MoS$_2$/MoS$_2$ structure.}
  \label{fig_graphene/mos2}
\end{figure}

\begin{figure}[htbp]  
  \begin{center}
    \scalebox{1.0}[1.0]{\includegraphics[scale=1]{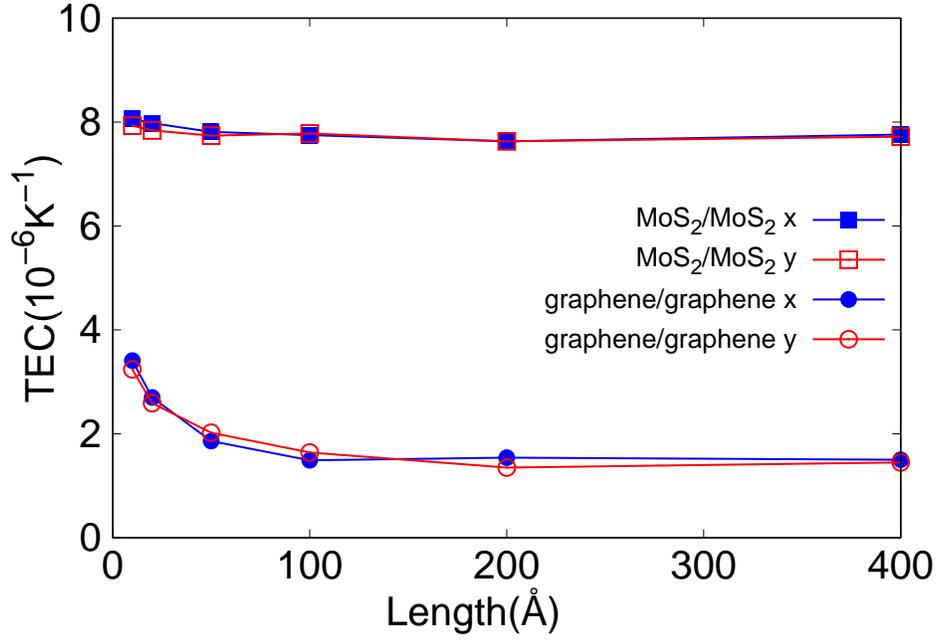}}
  \end{center}
  \caption{The size dependence for the TEC of graphene/graphene and MoS$_2$/MoS$_2$ structures, respectively.}
  \label{fig_size}
\end{figure}

\begin{figure}[htbp]  
  \begin{center}
    \scalebox{1.0}[1.0]{\includegraphics[scale=1]{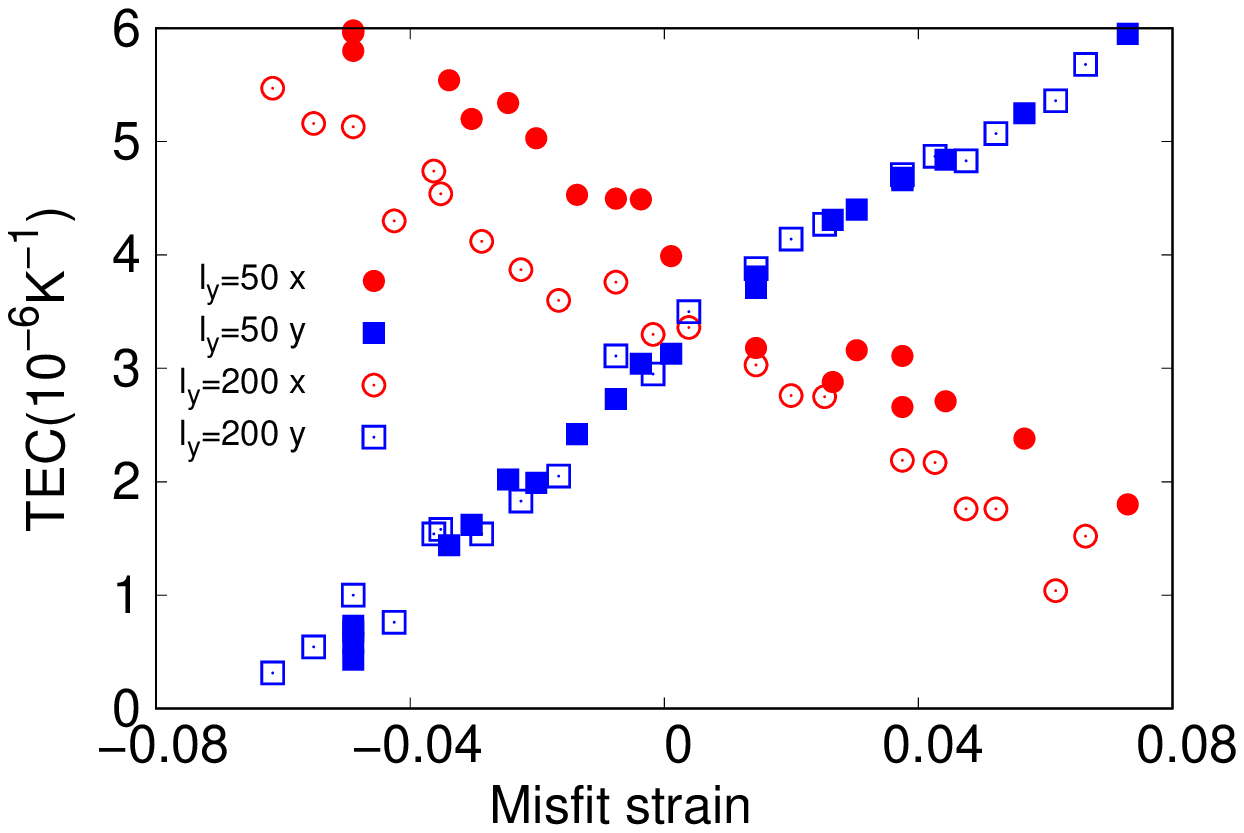}}
  \end{center}
  \caption{TEC for graphene/MoS$_2$ heterostructures with varying misfit strains.}
  \label{fig_strain-tec}
\end{figure}

Figure~\ref{fig_graphene/mos2} shows the temperature dependence for the sizes of graphene/graphene and MoS$_2$/MoS$_2$ bilayer structures. Panel (a) shows the temperature dependence for the two lateral dimensions for the graphene/graphene structure of dimension about $200\times 200$~{\AA}. The relation between the size and the temperature is fitted to a linear function, with a slope of $\frac{dL}{dT}$. The one-dimensional TEC, $\alpha$, is calculated by
\begin{eqnarray}
\alpha =\frac{1}{L_0}\frac{dL}{dT},
\label{eq_s}
\end{eqnarray}
where $L_0$ is the size at zero temperature. From Fig.\ref{fig_graphene/mos2}~(a), the TEC of the graphene/graphene structure is $1.54\times 10^{-6}$~{K$^{-1}$} and $1.35\times 10^{-6}$~{K$^{-1}$} along the x and y directions, respectively. We also calculate the TEC of monolayer graphene, which is a negative value of $-2.1\times 10^{-6}$~{K$^{-1}$} at 400~K. We find that the TEC of the graphene/graphene bilayer structure is larger than the monolayer graphene. It is because the negative TEC in monolayer graphene is mainly caused by the low-frequency flexural mode, which is limited in the graphene/graphene structure with larger bending modulus. Similar layer dependence for the TEC has also been found in previous Monte Carlo works.\cite{Zak2010Atomistic} The TEC for the MoS$_2$/MoS$_2$ structure is $7.63\times 10^{-6}$~{K$^{-1}$} along both x and y directions from Fig.\ref{fig_graphene/mos2}~(b). These value are consistent with the experiment (7.6 $\pm$ 0.9 $\times$ 10$^{-6}$~{K$^{-1}$} and 7.4 $\pm$ 0.5 $\times$ 10$^{-6}$~{K$^{-1}$})\cite{Zhang2019Thermal} and first-principle calculations (7.3 $\times$ 10$^{-6}$~{K$^{-1}$} and 7.2 $\times$ 10$^{-6}$~{K$^{-1}$})\cite{2015Effects,2014Correlation}. We find that the TEC is almost isotropic in both graphene/graphene and MoS$_2$/MoS$_2$ bilayer structures, which is as expected considering the three-fold rotational symmetry in both structures. 

Thermal induced ripples are important in two-dimensional materials like the graphene/graphene bilayer structure and the MoS$_2$/MoS$_2$ structure. As a result of the ripples, thermal properties in the atomic layers can be size dependent. Indeed, Fig.\ref{fig_size} shows that TEC depends on the size for these two bilayer structures. In particular, TEC is obviously larger in smaller structures. All structures simulated in this figure are of square shape with almost the same size in the two lateral dimensions. The size effect can be understood as follows. The heighth of the thermal induced ripple becomes larger at higher temperature, which consequently decreases the in-plane sizes for the bilayer structures. As a result, the ripple effect is to reduce the TEC. It should be noted that the thermal induced ripple has an intrinsic length around 30-60~{\AA} in graphene.\cite{2016Switchable} In graphene/graphene structures with dimension less than the intrinsic size, the thermal induced ripple is weakened due to the size confinement, so the negative effect from the ripple is weaker in smaller structures. Actually, Fig.\ref{fig_size} shows that the size effect becomes less important for larger graphene/graphene structures with size above 60~{\AA}, which further confirms the ripple based mechanism for the size effect. Hence, the TEC decreases with the increase of size for the graphene/graphene structure. Similar arguments are also applicable for the size dependence for the TEC of MoS$_2$/MoS$_2$ structures.

From the above, we have learned that MoS$_2$/MoS$_2$ structure has larger TEC than the graphene/graphene structure, and the thermal induced ripple has a negative effect on TEC. We now study the TEC for graphene/MoS$_2$ heterostructures. Fig.~\ref{fig_strain-tec} shows TEC for graphene/MoS$_2$ heterostructure structures with different misfit strain. Two sets of heterostructuers with quite different dimensions are simulated here. In the first set of heterostructures, the length along the y-direction is 199.2~{\AA} and 199.6~{\AA} for graphene and MoS$_2$ layers, respectively. These sizes are properly chosen, so that the misfit strain in the y-direction is minimum. The length along the x-direction for the graphene and MoS$_2$ layer is in the range of [187, 213]~{\AA}, so that the misfit strain is within [-6.2\%, 6.6\%]. In the second set of heterostructures, the length along the y-direction is 49.2~{\AA} and 49.9~{\AA} for graphene and MoS$_2$ layers, respectively. The length along the x-direction for the graphene and MoS$_2$ layer is in a wide range of [50, 105]~{\AA}.

The results in Fig.~\ref{fig_strain-tec} shows that the TEC can be linearly tuned from $1\times 10^{-6}$~{K$^{-1}$} to $6\times 10^{-6}$~{K$^{-1}$} by the misfit strain. More specifically, the TEC along the x-direction decreases linearly with increasing misfit strain, while the TEC along the y-direction increases linearly with the increase of the misfit strain. The effect of the misfit strain is quite robust, as similar phenomenon can be observed for both sets of heterostructures with quite different lateral dimensions.

\subsection{Misfit strain effect}

To reveal the underlying mechanism for the misfit strain effect, we derive an analytic formula for the relation between the TEC and the misfit strain. The total strain energy of a heterostructure related to the misfit strain is
\begin{eqnarray}
U =\frac{1}{2}E_1 \epsilon_1^2 + \frac{1}{2}E_2 \epsilon_2^2,
\label{eq_s1}
\end{eqnarray}
where $\epsilon_1=\frac{l-l_1}{l_1}$ and $\epsilon_2=\frac{l-l_2}{l_2}$ are the strain for graphene and MoS$_2$ layers, respectively. Here $l_1$ and $l_2$ are the original lengths for graphene and MoS$_2$ layers, respectively. The $l$ is the final length of the heterostructure structure after relaxation of the misfit strain. The $E_1$ and $E_2$ are the in-plane Young's moduli of graphene and MoS$_2$ layers, respectively. The final length $l$ can be obtained by minimizing the total strain energy with respective to $l$,
\begin{eqnarray}
l=\frac{E_1 l_2 + E_2 l_1}{E_1\frac{l_2}{l_1} + E_2\frac{l_1}{l_2}}.
\label{eq_l}
\end{eqnarray}
For a given temperature increasement of $\Delta T$, the graphene and MoS$_2$ layers are expanded to $l_1^{'} = l_1(1 + \alpha_1\Delta T)$  and $l_2^{'} = l_2(1 + \alpha_2\Delta T)$, respectively, where $\alpha_1$ and $\alpha_2$ are the TEC of graphene and MoS$_2$ layers. The length for the heterosructure is, 
\begin{eqnarray}
l'=\frac{E_1 l_2(1+\alpha_2\Delta T) + E_2 l_1(1+\alpha_1\Delta T)}{E_1\frac{l_2(1+\alpha_2\Delta T)}{l_1(1+\alpha_1\Delta T)} + E_2\frac{l_1(1+\alpha_1\Delta T)}{l_2(1+\alpha_2\Delta T)}}.
\label{eq_l'}
\end{eqnarray}
With the definition of TEC, $\alpha =\frac{l'-l}{l\Delta T}$, the following analytic expression is obtained for the TEC of graphene/MoS$_2$ van der Waals heterostructrues,
\begin{eqnarray}
\alpha =\frac{3E_1E_2(\alpha_1-\alpha_2)}{(E_1 + E_2)^2}\epsilon_m + \frac{\alpha_1E_1 + \alpha_2E_2}{E_1 + E_2}.
\label{eq_alpha-final}
\end{eqnarray}

Using the value of the Young's modulus and the TEC for graphene and MoS$_2$, we obtain the explicit relation between the TEC (in the unit of $10^{-6}$~{K$^{-1}$}) and misfit strain for the graphene/MoS$_2$ heterostructure
\begin{eqnarray}
\alpha =-3.65 \epsilon_m + 3.22.
\label{eq_alpha-expression}
\end{eqnarray}
This equation shows that the TEC is about 3.22 $\times$ 10$^{-6}$ K$^{-1}$ for the graphene/MoS$_2$ heterostructure without misfit strain, which agrees quite well with the MD simulation results. Furthermore, Eq.~(\ref{eq_alpha-expression}) shows that the TEC decreases linearly with the increase of misfit strain in the graphene/MoS$_2$, which is qualitatively consistent with the MD results. However, the slope for the linear function is only -3.65, which is about one order smaller than the MD result of -37.5 in Fig.\ref{fig_strain-tec}. It implies that some important factors must have been lost in deriving Eq.~(\ref{eq_alpha-expression}).

\subsection{Thermal induced ripple effect}

\begin{figure}[htbp]  
  \begin{center}
    \scalebox{1.0}[1.0]{\includegraphics[width=\textwidth]{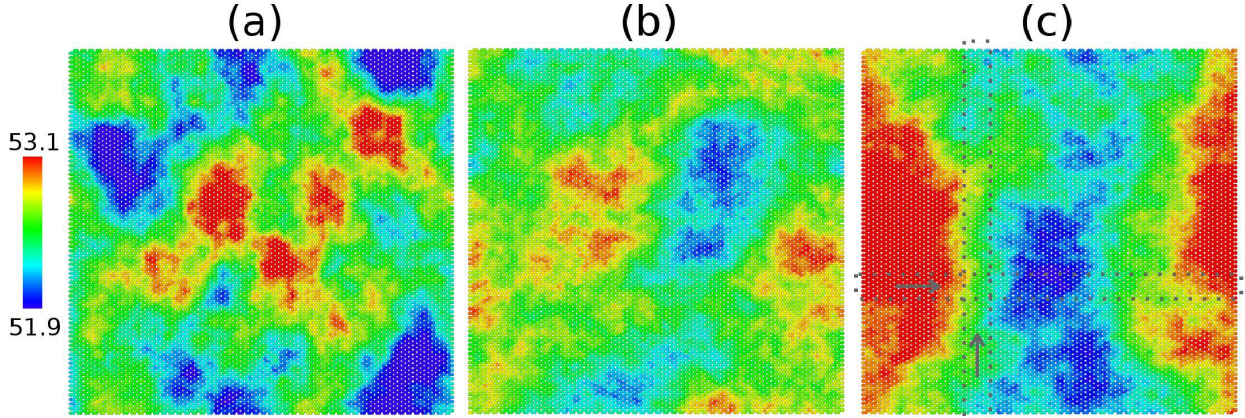}}
  \end{center}
  \caption{The thermal induced ripples in graphene/MoS$_2$ van der Waals heterostructures with three typical misfit strains: (a) $\epsilon_m=-6.2\%$, (b) $\epsilon_m=0.4\%$ and (c) $\epsilon_m=6.2\%$. Color bar represents the z-coordinate of each atom.}
  \label{fig_ripple}
\end{figure}

\begin{figure}[htbp]  
  \begin{center}
    \scalebox{1.0}[1.0]{\includegraphics[scale=0.7]{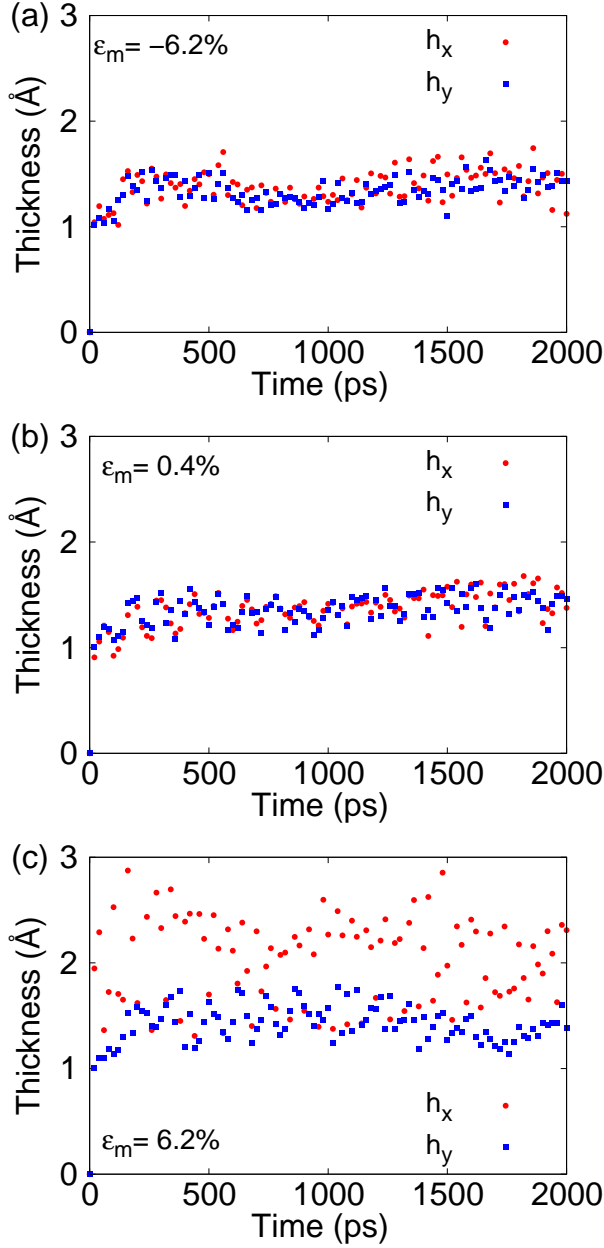}}
  \end{center}
  \caption{The time history for the effective thickness of the graphene/MoS$_2$ with three typical misfit strains: (a) $\epsilon_m=-6.2\%$, (b) $\epsilon_m=0.4\%$, and (c) $\epsilon_m=6.2\%$. The temperature is 300~K.}
  \label{fig_hx-time}
\end{figure}

\begin{figure}[htbp]  
  \begin{center}
    \scalebox{1.0}[1.0]{\includegraphics[scale=0.7]{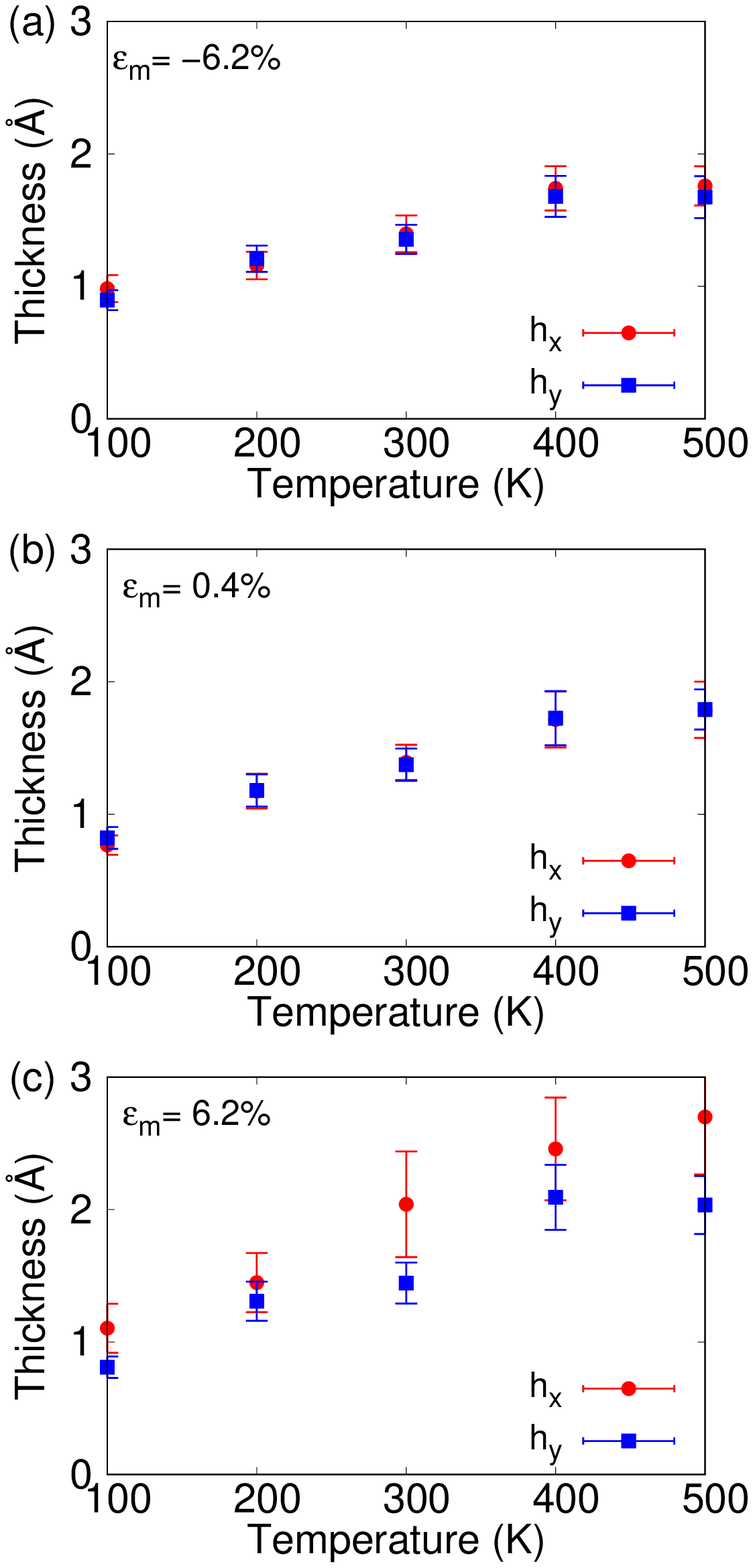}}
  \end{center}
  \caption{The temperature dependence for the averaged effective thickness of the graphene/MoS$_2$ with three typical misfit strains: (a) $\epsilon_m=-6.2\%$, (b) $\epsilon_m=0.4\%$, and (c) $\epsilon_m=6.2\%$.}
  \label{fig_hx-temp}
\end{figure}

It should be recalled that the thermal induced ripples play an important role in thermal and mechanical properties of the quasi-two-dimensional atomic layers like graphene/MoS$_2$ heterostructures. Particularly, the thermal induced ripple has negative effect on the TEC as shown in the above. Previous works have illustrated that the mechanical strain is an efficient way to tune the thermal induced ripples in atomic-layered structures\cite{Hussain2020Strain,Wang2017Origins,Kou2015Anisotropic,Fairs2020Ripples}. It is quite possible that the misfit strain (a specific mechanical strain) can affect the thermal induced ripples and may eventually influence the TEC of the graphene/MoS$_2$ heterostructure. We thus investigate the effect of the misfit on the thermal induced ripples in Fig.~\ref{fig_ripple}, where three quite different misfit strains are considered. The misfit strain is -6.2\%, 0.4\%, and 6.2\% for these three structures. Obviously, the structure with large misfit strain in panel (c) has anisotropic ripples along the two in-plane directions, where the structure is slightly buckled along the horizontal x-direction.

The misfit strain effect on the ripples can be further quantified by introducing the effective thickness along the x and y directions, i.e., $h_x$ and $h_y$. We take Fig.~\ref{fig_ripple}~(c) as an example to illustrate the definition and calculation of $h_x$. Scanning the figure from left to right along the horizontal x-direction, one gets the minimum and maximum value of the z-coordinate, i.e., $z_{\rm min}$ and $z_{\rm max}$. The thickness along the x-direction is $h_x=z_{\rm max}-z_{\rm min}$. Similarly, $h_y$ is calculated by scanning along the vertical y-direction. The thickness for Fig.~\ref{fig_ripple}~(c) is $h_x=1.10$~{\AA} and $h_y=0.81$~{\AA}. The effective thickness is anisotropic along these two in-plane directions, which truly represents the major feature of this structure.

Figure~\ref{fig_hx-time} shows the time history for the effective thickness at room temperature. Panels (a) and (b) show that the thickness is isotropic along the x and y directions for structures with $\epsilon_m=$ -6.2\% and 0.4\%. Panel (c) shows that the graphene/MoS$_2$ heterostructure with large misfit strain of 6.2\% has a large effective thickness along the x-direction. The effective thickness can be obtained by averaging over time. For example, we get $h_x=1.10 \pm 0.18$~{\AA} and $h_y=0.81 \pm 0.08$~{\AA} for the structure in Fig.~\ref{fig_ripple}~(c), where the error bar is the standard deviation error in the average over time.

\begin{figure}[htbp]  
  \begin{center}
    \scalebox{1.0}[1.0]{\includegraphics[scale=1]{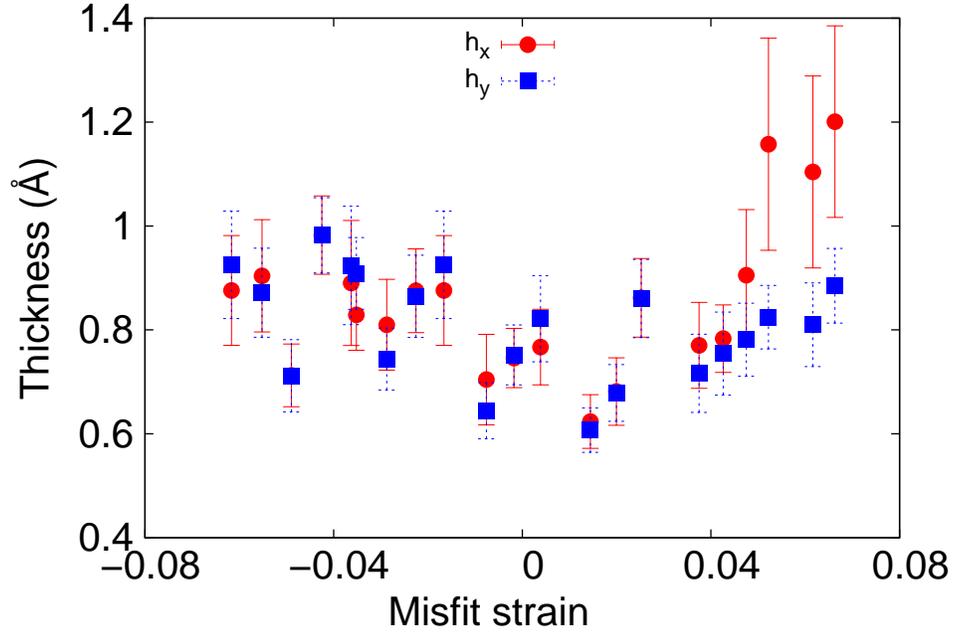}}
  \end{center}
  \caption{The effective thickness for the graphene/MoS$_2$ heterostructure of different misfit strain at 100~K.}
  \label{fig_hxall}
\end{figure}

\begin{figure}[htbp]  
  \begin{center}
    \scalebox{1.0}[1.0]{\includegraphics[scale=1]{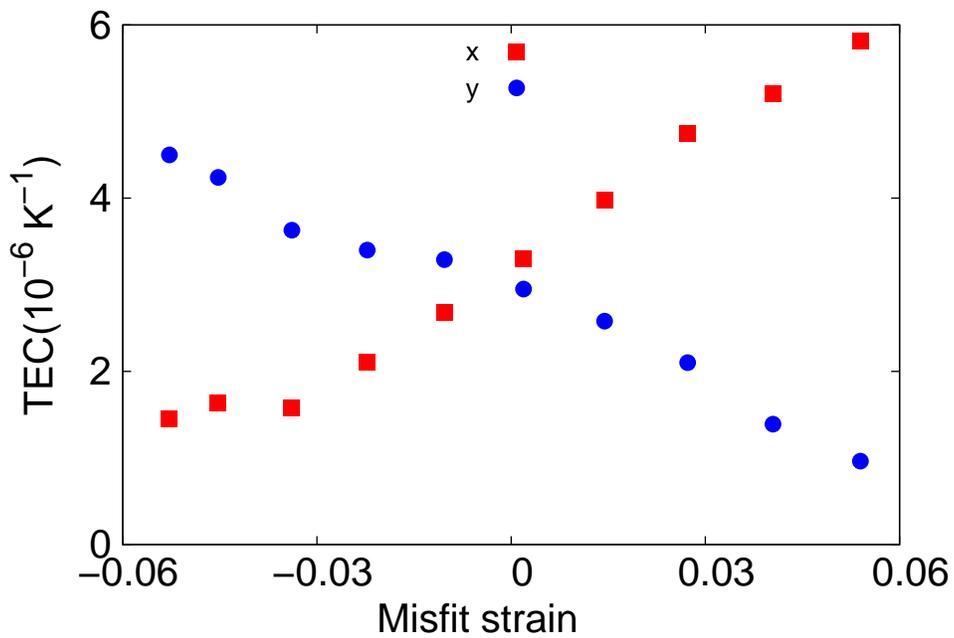}}
  \end{center}
  \caption{The TEC for graphene/MoS$_2$ heterostructure with varying misfit strain along the y-direction. The misfit strain along the x-direction is almost zero (-0.2\%).}
  \label{fig_misy}
\end{figure}

The temperature dependence for the effective thickness is shown in Fig.~\ref{fig_hx-temp} for these three graphene/MoS$_2$ heterostructures with different misfit strain. Panel (c) shows that the effective thickness is anisotropic in the structure with large misfit strain of 6.2\%. Fig.~\ref{fig_hxall} further illustrates that the effective thickness is anisotropic for structures with the misfit strain above 4\%, where $h_x$ is obviously larger than $h_y$. It is because the graphene layer is stretched while the MoS$_2$ layer is compressed by the misfit strain in the graphene/MoS$_2$ heterostructure with positive misfit strain. The MoS$_2$ layer is seriously compressed and will be slightly buckled as intrigued by the thermal vibration. The buckling of the MoS$_2$ layer will lead to the buckling of the whole heterostructure, as the graphene layer is much softer than the MoS$_2$ layer. The large thermal induced ripple along the x-direction causes a strong reduction of the TEC along the x-direction, leading to the decrease of the TEC along the x-direction with the increase of the misfit strain in Fig.\ref{fig_strain-tec}. In contrast, the buckling along the x-direction can weaken the thermal induced ripples along the y-direction, so the y-directional TEC increases with increasing misfit strain. We have thus explained misfit strain effect on the TEC for $\epsilon_m>0$ based on the thermal induced ripples.

For $\epsilon_m<0$, the graphene layer is compressed while the MoS$_2$ layer is stretched by the misfit strain. Although the graphene layer is compressed seriously, the whole heterostructure is not buckled as can be seen from Fig.~\ref{fig_ripple}~(a), Fig.~\ref{fig_hx-time}~(a), and Fig.~\ref{fig_hx-temp}~(a). The situation becomes more complex, as several competing mechanisms co-exist. On the one hand, the compressive misfit strain on the graphene layer tries to reduce the TEC along the x-direction, because compression can induce ripples that have negative effect on TEC. On the other hand, the positive misfit strain on the MoS$_2$ layer tries to increase the TEC along the x-direction, because the nonlinear effect becomes stronger upon stretching. The nonlinear effect will enhance the TEC. Our MD results show that the second mechanism surpasses the first mechanism, so the TEC along the x-direction increases with decreasing misfit strain for $\epsilon_m<0$. One possible reason is that the TEC for MoS$_2$ is much larger than the TEC for graphene as can be seen from Fig.~\ref{fig_size}, so the second mechanism (on MoS$_2$) is more important than the first mechanism (on graphene).

In all above graphene/MoS$_2$ heterostructures, the misfit strain is almost zero along the y-direction, while the misfit strain in the x-direction is changed to tune the TEC. Fig.~\ref{fig_misy} shows similar misfit strain effect on the TEC in graphene/MoS$_2$ heterostructures with varying misfit strain along the y-direction.

\section{Conclusion}

To summarize, the present work studies the effect of the misfit strain on the TEC of the graphene/MoS$_2$ heterostructure by MD simulations. The TEC can be tuned in a wide range from $1\times 10^{-6}$~{K$^{-1}$} to $6\times 10^{-6}$~{K$^{-1}$} by the misfit strain, and this misfit strain effect is not sensitive to the size or direction of the heterostructure. More specifically, the TEC along the misfit strain direction decreases linearly with the increase of misfit strain. Several possible underlying mechanisms are explored to be responsible for the misfit strain effect on the TEC. We first derive an analytic formula to directly relate TEC to the misfit strain, which qualitatively agrees with the MD results, but this analytic formula underestimates the misfit strain effect. We further demonstrate that the thermal induced ripples can be tuned by the misfit strain, which in turn causes strong effects on the TEC of the heterostructure. These results provide an effective manner to manipulate the TEC of van der Waals heterostructures through the misfit strain engineering.

\textbf{Acknowledgment} The work is supported by the National Natural Science Foundation of China (Grant Nos. 11822206 and 12072182) and Innovation Program of the Shanghai Municipal Education Commission (Grant No. 2017-01-07-00-09-E00019).

\end{document}